# Establishing Digital Recognition and Identification of Microscopic Objects for Implementation of Artificial Intelligence (AI) Guided Microassembly

Tuo Zhou[1,2], Shih-Yuan Yu[3], Matthew Michaels[1,2], Fangzhou Du[3], Lawrence Kulinsky[1], Mohammad Abdullah Al Faruque[1,3]

[1] Mechanical and Aerospace Engineering, University of California Irvine, 5200 Engineering Hall, Irvine, CA 92627, USA

[2] Materials and Manufacturing Technology, University of California Irvine, 5200 Engineering Hall, Irvine, CA 92627, USA

[3] Electrical Engineering and Computer Science, University of California Irvine, 2200 Engineering Hall, Irvine, CA 92627, USA

**Abstract**

As miniaturization of electrical and mechanical components used in modern technology progresses, there is an increasing need for high-throughput and low-cost micro-scale assembly techniques. Many current micro-assembly methods are serial in nature, resulting in unfeasibly low throughput. Additionally, the need for increasingly smaller tools to pick and place individual microparts makes these methods cost prohibitive. Alternatively, parallel self-assembly or directed-assembly techniques can be employed by utilizing forces dominant at the micro and nano scales such as electro-kinetic, thermal, and capillary forces. However, these forces are governed by complex equations and often act on microparts simultaneously and competitively, making modeling and simulation difficult. The research in this paper presents a novel phenomenological approach to directed micro-assembly through the use of artificial intelligence to correlate micro-particle movement via dielectrophoretic and electro-osmotic forces in response to varying frequency of an applied non-uniform electric field. This research serves as a proof of concept of the application of artificial intelligence to create high yield low-cost micro-assembly techniques, which will prove useful in a variety of fields including micro-electrical-mechanical systems (MEMS), biotechnology, and tissue engineering.

**Keywords:** Micro- and Nano- Assembly, Artificial Intelligence, Directed-Micro-Assembly

## 1. Introduction

Miniaturization of electrical and mechanical systems has fuelled a dramatic increase in the reliance of modern technology on integrated circuits and micro-electro-mechanical systems (MEMS). As such, research into manufacturing and assembly at the micro and nano scale is of increasing interest and necessity. The goal of micro-assembly is to create multi-part micro-scale devices of high complexity with high yield and low cost [1].

Many current micro-assembly methods are serial processes, in which micro parts are assembled one at a time. A common method for serial micro-assembly is the use of micro-grippers to pick and place individual microparts. This method of assembly suffers from being too slow to achieve industrially feasible throughput. Additionally, the tools required to perform serial micro-assembly must become increasing smaller to obtain better assembly resolution, an expensive and slow proposition.

An alternative method for micro-assembly is guided, or directed, assembly in which many microparts can be assembled in parallel. One example of such techniques is the assembly of inorganic microparts via polymer-guided assembly [2]. Commonly, methods for directed assembly utilize forces dominant on the micro scale such as electrostatics, surface and capillary forces, and thermal forces. The present study relies primarily on Dielectrophoretic (DEP) and Electro-osmotic (EO) forces. DEP describes a force acting on the induced dipole moment of a particle suspended in dielectric fluid resulting from an applied non-uniform electric field [3]. EO refers to the movement of fluid resulting from ion build-up at a charged surface within an applied electric field [4].

A major issue with directed micro-assembly is the complexity of the forces acting in the micro-domain. Many of these forces such as DEP and electro-osmosis act simultaneously on the microparts and thus successful prediction of the behaviour of microparts is often difficult to achieve. In the absence of sufficient isolation of a single force or comprehensive model of the various forces acting on the microparts, artificial intelligence (AI) can be used to evaluate the response of microparts to the applied signals (frequency and voltage for DEP) and to guide the manipulation of these parts.

Specifically, present study is looking at the utilization of algorithmic AI, where the behaviour of microparticles is first captured via digital camera and then evaluated mathematically, by comparison of the position of particles in individual frames. The mathematical algorithms are programmed into the software to quantify the output (determination if particles are moving towards or away from the electrodes) and then correlate this behavior of the particles to the given input variables such as the frequency of the applied electric field.

In this study, a novel phenomenological approach to micro-assembly utilizing AI is presented. A closed-loop cyber-physical system was developed to characterize the response of polystyrene micro-beads to changes in input of an applied electric field

(including voltage and the applied frequency). Digital camera captures a sequence of images that are digitized. Image processing is used to recognize microparts and determine their pattern of movement – towards the electrodes or away from the electrodes. Having determined the type of the bead movement under the specific input conditions (for example, given frequency), the program changes the input conditions, and the new sequence of images is analysed. Therefore, the system is capable of determining frequency ranges in which the beads are attracted to or are repelled from the electrodes as well as ranges in which the beads are unaffected by the electric field.

The phenomenological approach to directed micro-assembly presented in this paper will find application across a variety of fields including microsystems and electronics, biotechnology, drug delivery, and tissue engineering [5].

## 2. Methods and Materials

### 2.1. Fabrication of Electrodes

The interdigitated gold electrodes (IDEs) were fabricated via photolithography and e-beam evaporation process. Initially, a thin layer of positive photoresist (Shipley) was spin-coated onto a 4-inch silicon wafer covered with a 1 µm thick thermal oxide layer (University Wafer, South Boston, MA, USA) using a Laurell photoresist spinner (Laurell Technologies, North Wales, PA, USA) at an initial speed of 3000 rpm for 30s. Then, the resist was soft-baked at 90 °C for 30 minutes on a hot plate (Dataplate, Pmc, 732 Series, Dubuque, IA, USA). Next, the resist layer was exposed through a photomask (CadArt, Bandon, OR, USA) to a UV light source at an energy intensity of $10 mW/cm^2$ for 35s using the Karl Suss MA56 Mask Aligner ( Karl Suss, Garching, Germany). The portions of the resist layer exposed to the UV light were washed away using deionized water.

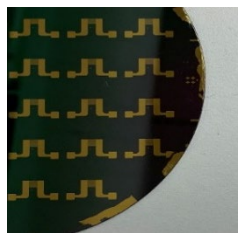

Fig. 1. Fabricated gold IDEs in 6" silicon wafer.

A 300 Å layer of chromium was deposited onto the wafer using a Temescal CV-8 e-beam evaporator (AIRCO.INC, Berkeley, CA, USA), followed by the deposition of a second layer (300 nm thick) – that of gold. The remaining photoresist was washed away by acetone to leave the set of gold IDEs (see Fig. 1). Each IDE consisted of 12 individual fingers 70 µm in width,

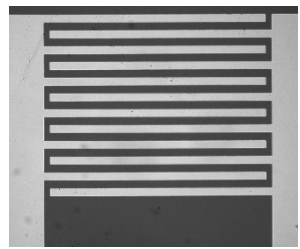

Fig. 2. Electrode fingers of the IDEs. Each electrode finger is 70 microns wide. The interdigitated electrodes are located on the top of each Π shaped electrode set seen on Figure 1.

separated by a gap of 70 µm (see Fig. 2).

### 2.2 Experimental Setup

34-gauge buss wires (Guasti Wire and Cable, Ontario, Canada) were soldered to both contact pads of the IDEs using indium solder. The chip containing the IDEs was then placed under an optical microscope (Nikon Eclipse, Minato, Japan) connected to a cMOS digital camera (SPOT Imaging, Sterling Heights, MI, USA). Each buss wire was then connected to a Siglent 2082X function generator (Siglent Technologies, Solon, OH, USA) to induce an electric field across the fingers of the IDE. 5 µL of polystyrene bead suspension, containing 3 µm diameter beads, were pipetted onto the IDEs and a slide cover was laid on top to reduce evaporation.

### 2.3 Hardware/Software Integration

The camera and function generator were integrated into the software program to create the closed loop system. The live images of the beads were captured using the PyAutoGUI Python package and was processed using OpenCV. The function generator was connected via USB to the computer and the frequency of the applied electric field was controlled via SCPI commands using the PyVISA Python package.

### 2.4 Software Architecture

The closed-loop system analyses and directs the movement of the polystyrene beads using information from the cMOS camera enabled optical microscope to monitor the gaps between the IDE fingers (see Fig. 3). Designed as a real-time embedded system, this system constantly monitors the testbed and makes decisions on adjusting the frequency of the electric field applied across the testbed which in turn changes the magnitude and direction of the DEP force experienced by the polystyrene beads.

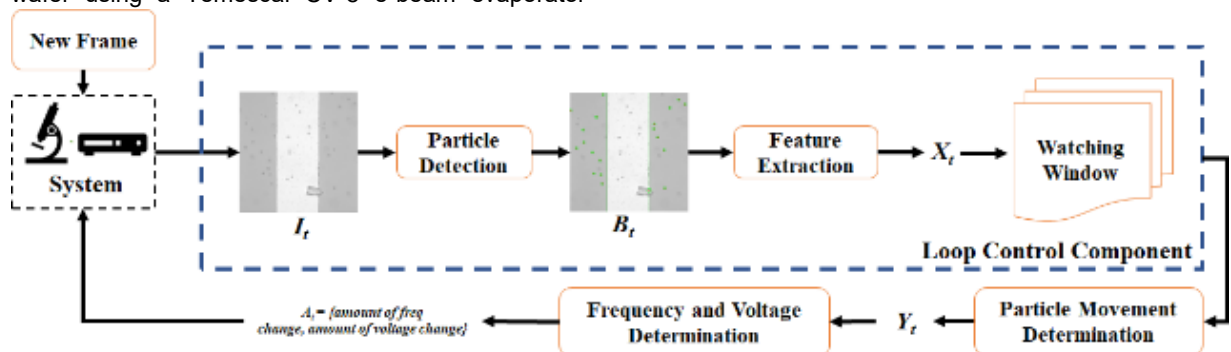

Fig. 3. Software Architecture

*2.4.1 Particle Detection*

For each frame $I_t \in I$, the system utilizes *Hough Circle Detector* (HCD) to detect particles [6]. The HCD leverages the *Hough Gradient Method* implementation in OpenCV [7] that detects the circle-shaped objects on images [6] (see Fig. 4). For ensuring the detecting performance across various experimental setups, the system has four parameters for tuning HCD's performance: *param_1*, *param_2*, *min_radius*, *max_radius*. Specifically, *param_1 is related to* the internal Canny detector threshold and *param_2* is related to the center detection threshold. For each $I_t$, the detection result is denoted as a container of particles $B_t$. Such process is abstracted as a converting function $F$. Thus, the relation between a new frame $I_t$ and a container $B_t$ can be described as: $F(I_t) = B_t = \{P_1, P_2, \ldots P_N\}$ where $N$ is the number of detected particles and each particle is represented with the center 2-D coordinates (i.e. $P_n = \{x_n, y_n\}$).

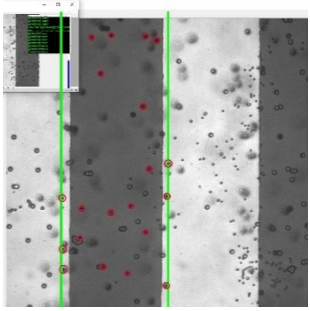

Fig. 4. Particle Detection via HCD. Green lines indicate visual recognition of electrodes' edges and constitute the observation window.

*2.4.2 Feature Extraction*

Once detected, the system then transforms $B_t$ into a corresponding feature vector, denoted as $X_t$. Two assumptions are made. The first one is that the particle movement can be represented by the values derived from formulaic methods. Second, the direction of DEP force is assumed to be perpendicular to the electrodes of the testbed. Therefore, the system considers a particle's movement along the x-axis. The process of acquiring $X_t$ is abstracted as another converting function $G$, realized by calculating the average absolute distance to the reference line where the reference line refers to a vertical line that stays in between and in the middle of two electrodes. The relationship between a new frame $I_t$ to the extracted feature $X_t$ is formulated as:

$$X_t = G(F(I_t)) = G(B_t) = \sum_{n=1}^{N} \frac{|x_n - r|}{N}$$

where $r$ denotes the x-coordinate of the reference line.

*2.4.3 Movement Determination*

To capture the macroscopic motion of particles at each timestamp t, the system performs a linear trend analysis by considering a subset of features between current frame and few frames prior. This subset is defined as $W = \{X_{t-k}, X_{t-k+1}, \ldots X_t\}$ where k denotes the length of $W$. Our system first post-processes the features in $W$ with *Missing-Value Sampling* and *Data Smoothing*. As HCD might end up not detecting any particle for $I_t$, the function in *Missing-Value Sampling* finds an alternative value of $X_t$, thus minimizing the negative influence of missing values in the linear trend analysis. Besides, the system runs *Data Smoothing* on $W$ so that the influence of noise or random errors could be minimized. Specifically, a linear convolution with an unweighted filter is applied on $W$, allowing the system to acquire a cleaner trend during the analysis.

Next, the proposed system uses *Linear Trend Model* (LTM) for determining the macroscopic motion of the detected particles. To be more specific, the system performs the linear regression to derive a trend with all the features maintained in $W$. For all features in $\{X_{t-k}, X_{t-k+1}, \ldots X_t\}$, the least-squares regression line generates a unique trend line represented by equation y = bx + c that minimizes the vertical distance from each $X_t$ to the regression line, and the coefficient, $b$, is calculated as: $b = \frac{\sum(x-\bar{x})*(y-\bar{y})}{\sum(x-\bar{x})^2}$

where the coefficient $b$ represents the collective velocity of all particles. The system uses $b$ and a decision threshold $\delta$ to classify the macroscopic motion of particles with a categorical label $Y_t$. If $|b| \leq \delta$, the system classifies the motion as *NO_DEP* because as $b$ is too little to be considered as a DEP polarity. In the case where $|b| > \delta$, the system classifies the DEP's polarity as either Positive-DEP or Negative-DEP.

*2.4.4 Feedback Control Design*

The proposed system implements AI-guidance using cyber-physical feedback control system. With a predefined sampling rate m, the system repeatedly collects and analyses the new frame from the camera to acquire a feature $X_t$ as described in Sec. 2.4.1 ~ 2.4.3. In runtime, each new feature $X_t$ is inserted to the end of the watching window W which is realized as *FIFO* (First-In First Out) Queue with k as the size in the system. Next, in *Non-SETTLE* state, the system computes $Y_t$ that indicates the current testbed state from the watching window $W = \{X_{t-k}, X_{t-k+1}, \ldots X_t\}$. In *SETTLE* state, the system keeps obtaining inputs from the camera and extracting $X_t$.

According to $Y_t$ and $b$, the system adjusts the frequency or voltage applied by the function generator to the electrodes. The adjustments are encapsulated as a command packet and sent to the function generator. Once the applied signal is adjusted, the system switches its state to *SETTLE*. In *SETTLE*, the system detects the change of the microscopic motion for particles induced by the function generator. The system monitors two changes: *particle response time* and *system response time*. The *particle response time* refers to the time required for particles to manifest the effect of function generator's manipulation. The *system response time* refers to the time needed by the system's analysing components to capture the macroscopic movement of particles.

## 3. Results and Discussion

The cyber-physical system detailed in section 2 above is allowed to run for a total of 4 cycles (applied signal via function generator, image acquisition, image analysis, data processing/algorithmic guidance, change is signal implemented by the function generator). The system demonstrated capability to accurately detect the polystyrene beads, determine their movement resulting from DEP and EO forces, and adjust the frequency of the electric field accordingly. On average, the Hough Circle Detector was capable of identifying 20-25% of the beads in each frame, a suitable sample size to estimate the overall movement of the beads.

During the execution of the program, 4 distinct frequencies were tested and the average absolute distance of the detected particles from the center of the testbed was plotted as a function of frame number, and by extension, time (see Fig. 5). The plot shows a clear increase in average absolute distance when frequency

## 4. Conclusions

The cyber-physical system detailed in this study successfully demonstrated the capability to identify micro particles, algorithmically determine their movement resulting from DEP and/or EO forces and correlate this movement to changes in the frequency of the applied electric field.

These results serve as a proof of concept that artificial intelligence can be applied to establish a phenomenological approach to directed micro-assembly. By utilizing micro-domain forces such as DEP and EO a parallel assembly is achieved which could offer significantly higher throughput compared to existing serial assembly techniques. Additionally, because the magnitude of DEP force is dependent on the size of the particles [8], the system and experimental setup presented in this research offers potential for studying selectivity in heterogeneous systems with particles of varying sizes and within various media. This initial research design will serve as a building block for further research into application of artificial intelligence for micro and nano assembly.

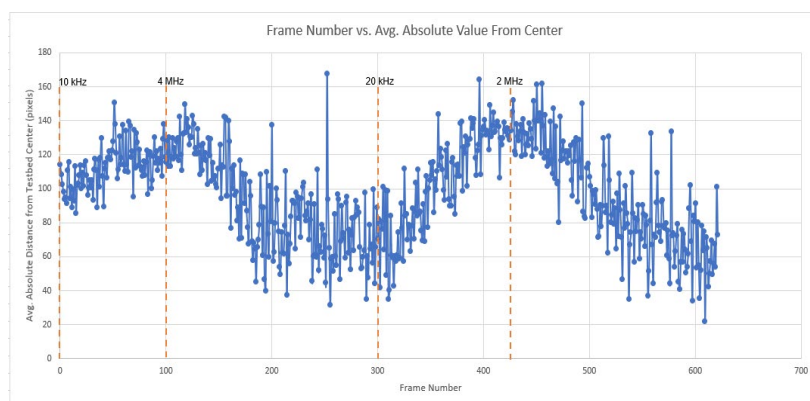

Fig. 5. Plot of average absolute distance from testbed center as a function of frame number

was set to 10 kHz and 20 kHz, correctly identifying the positive DEP and EO force attracting the beads to the surface of the electrodes at low frequencies. Conversely when frequency was set to 4MHz and 2MHz, the plot shows a decrease in average absolute distance, indicating the negative DEP force expected at high frequency.

Figures 6a and 6b contain images of single frames taken by the program at various points during testing. Frame number 300 (see Fig. 6a) was taken while the frequency was set to 4 MHz and the frame illustrates the beads being attracted to the surface of the electrodes via positive DEP and EO. Frame number 414 (see Fig. 6b) was taken while the frequency was set to 20 kHz and illustrates the beads being repelled from the electrodes towards the center of the testbed.

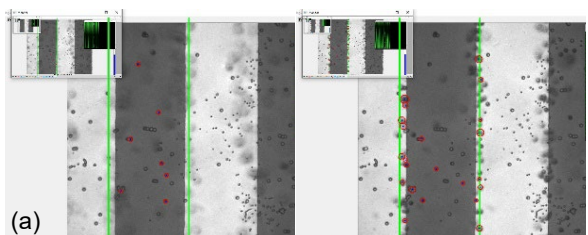

Fig. 6. (a) Frame 300 at 4 MHz (b) Frame 414 at 20 kHz.

**Acknowledgements**
The authors acknowledge the support of the National Science Foundation (award CMMI-1661877).

**References**
[1] K.F. Borhringer et al., Handbook of Industrial Robotics, John Wiley & Sons, 1999
[2] C. Yi et al., "Polymer-Guided Assembly of Inorganic Nanoparticles", Chem. Soc. Reviews, 2020; 40: 465-508.
[3] K. Kendall, "Electromechanics of Particles", Powder Tech., 1996; 89: 177-178.
[4] V.K. Narla, et al., "Electroosmosis Modulated Transient Blood Flow in Curved Microvessels, Study of a Mathematical Model", Microvascular Res., 2019; 123: 25-34.
[5] Y. Du et al., "Surface-Directed Assembly of Cell-Laden Microgels", Biotech. And Bioeng., 2010; 105: 655-662
[6] H.K. Yuen et al., "Comparative Study of Hough Transform Method for Circle Finding", Image and Vision Computing, 1990; 8: 71-77
[7] G. Bradski, "The OpenCV Library", Dr. Dobb's Journal of Software Tools, 2008; 19: 21-54.
[8] T. Zhou et al., "Guided Healing of Damaged Microelectrodes via Electrokinetic Assembly of Conductive Carbon Nanotube Bridges", Micromachines, 2021; 12: 405.